# A Method for Expressing and Displaying the Vehicle Behavior Distribution in Maintenance Work Zones

Qun Yang [1], Zhepu Xu [1], Saravanan Gurupackiam [2], Ping Wang [1*]

[1] Key Laboratory of Road and Traffic Engineering of Ministry of Education, Tongji University, Shanghai, China, 201804
[2] Department of Civil Engineering, Pennsylvania State University, Harrisburg, Middletown, PA, USA, 17057
* lilyshwang@tongji.edu.cn

## Abstract

Maintenance work zones on the road network have impacts on the normal travelling of vehicles, which increase the risk of traffic accidents. The traffic characteristic analysis in maintenance work zones is a basis for maintenance work zone related research such as layout design, traffic control and safety assessment. Due to the difficulty in vehicle microscopic behaviour data acquisition, traditional traffic characteristic analysis mainly focuses on macroscopic characteristics. With the development of data acquisition technology, it becomes much easier to obtain a large amount of microscopic behaviour data nowadays, which lays a good foundation for analysing the traffic characteristics from a new point of view. This paper puts forward a method for expressing and displaying the vehicle behaviour distribution in maintenance work zones. Using portable vehicle microscopic behaviour data acquisition devices, lots of data can be obtained. Based on this data, an endpoint detection technology is used to automatically extract the segments in behaviour data with violent fluctuations, which are segments where vehicles take behaviours such as acceleration or turning. Using the support vector machine classification method, the specific types of behaviours of the segments extracted can be identified, and together with a data combination method, a total of ten types of behaviours can be identified. Then the kernel density analysis is used to cluster different types of behaviours of all passing vehicles to show the distribution on maps. By this method, how vehicles travel through maintenance work zones, and how different vehicle behaviours distribute in maintenance work zones can be displayed intuitively on maps, which is a novel traffic characteristic and can shed light to maintenance work zone related researches such as safety assessment and design method.

## 1 Introduction

Maintenance work zones gain much attention because their existences change the vehicle trajectories, reduce the travel stability, cause congestion and increase the risk of traffic accidents. How vehicles travel in maintenance work zones, how the layouts of work zones affect the travel of vehicles, and whether the maintenance work zones are safely placed are issues of great interest to researchers.

The traffic characteristic analysis in maintenance work zones is a basis for the researches mentioned above. Due to the difficulty in gathering vehicle microscopic behaviour data such as acceleration, traditional researches are mainly limited to macroscopic traffic characteristics like average speed, distance headway and volume [1-4]. To gain a deeper understanding of the traffic characteristics in maintenance work zones and further improve the work related to maintenance work zones, there is a need to study from the point of view of the microscopic level. With the development of data acquisition technology, this kind of microscopic research becomes possible.

In this paper, portable vehicle microscopic behaviour data collection devices (PVMBDADs) are used, by which a large amount of microscopic behaviour data can be obtained. Based on the data, the behaviours of vehicles travelling through work zones are finely expressed and the distributions of vehicle behaviours are intuitively displayed on maps, which can show us a new point of view towards the traffic characteristics in maintenance work zones.

Vehicle behaviour refers to the reaction of the vehicle after the action of the driver and environment, such as linear accelerating, linear decelerating, and turning, et al. The goal of expressing vehicle behaviour is to characterise and distinguish different behaviours. As for an actual driving process, it is composed of several basic behaviour periods (stopping, linear accelerating, turning, et al.), as shown in Fig. 1. One main work of this study is to express the basic vehicle behaviours. To be more specific, the following 10 kinds of vehicle behaviours are expressed in this research: (1) stopping, (2) straight line driving with constant speed (L&C), (3) linear accelerating (L&A), (4) linear decelerating (L&D), (5) turning left and accelerating (TL&A), (6) turning right and accelerating (TR&A),(7) turning left with constant speed (TL&C), (8) turning right with constant speed (TR&C), (9) turning left and decelerating (TL&D) and (10) turning right and decelerating (TR&D). Fig. 2 shows the different behaviours defined above.

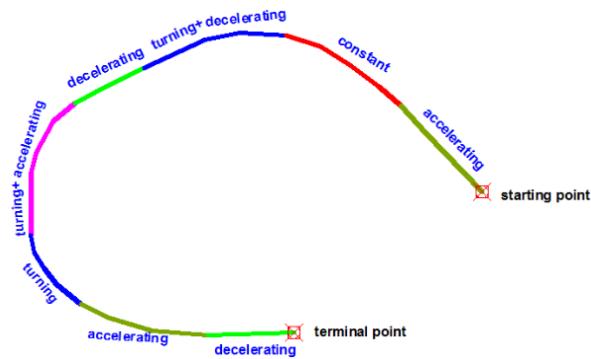

**Fig. 1.** The relationship between basic periods and a whole process

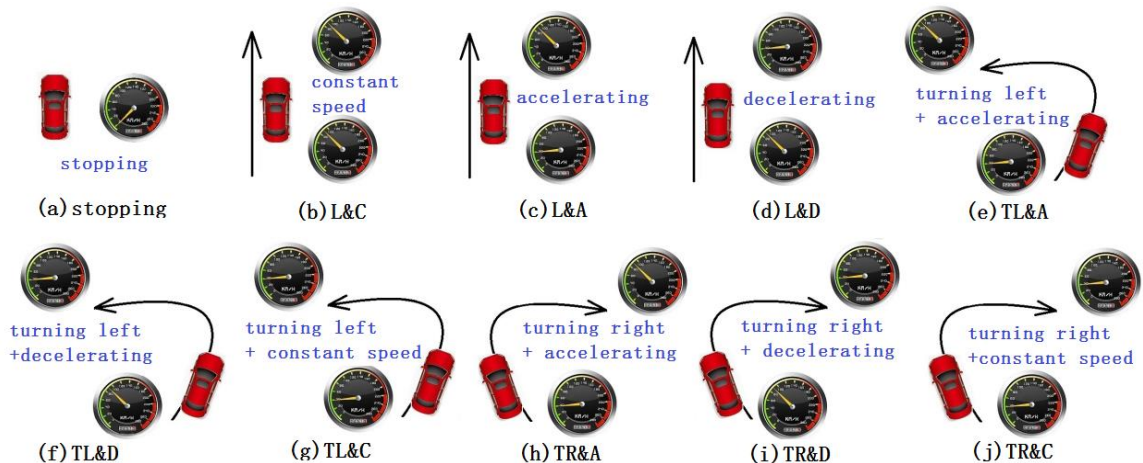

**Fig. 2.** 10ten basic vehicle behaviours

There are lots of vehicles running on the road network every day, under the impacts of maintenance work zones, vehicles behaviours will show a common pattern in space, such pattern is called vehicle behaviour distribution in this paper. For example, when travelling through work zones, maybe in the close upstream of the maintenance work zone most vehicles will decelerate there. The behaviour distribution is also an important characteristic of traffic flow. Studying the behaviour distribution characteristic will shed new light to maintenance work zone related researches.

The remaining part is organised as follows: Section 2 reviews some work related to this research from two aspects, i.e. the traffic characteristic analysis in maintenance work zones and the vehicle behaviour expressing method. Section 3 gives a brief introduction to the data acquisition device and method used in this research. The vehicle behaviour expressing and distribution displaying methods are established in the fourth and fifth section, respectively. Section 6 provides a summary and future course of this study.

## 2 Related work

### 2.1 Traffic characteristic analysis in maintenance work zones

Traffic characteristics analysis is a basis for important work such as layout design, traffic control and safety assessment in maintenance work zones. In the past, due to the limitation of insufficient data collection tools, collecting traffic data in maintenance work zones requires manual work and causes errors. Collectors can only make a single point measurement to obtain the traffic volume and distance headway of the observation point. The speed collecting is achieved by calculating the average speed of a short road segment to approximate the speed at that section of the road. Therefore, the researches on the traffic characteristics of maintenance work zones were mainly given to the characteristics of macroscopic traffic flow, analysing the vehicle running speed, traffic capacity, queuing and delay [1-4]. However, all kinds of complex macroscopic traffic phenomena are caused by the interactions between individual vehicles. At present, we have a very limited understanding of the behaviours of individual vehicles in maintenance work zones [5]. To gain a deeper understanding of the traffic characteristics in maintenance work zones and further improve the work related to maintenance work zones, there is a need to move from a macroscopic level to a microscopic level. With the development of data acquisition technology, this kind of microscopic research becomes possible.

Researchers at the Saxton Transportation Operations Laboratory at FHWA's Turner-Fairbank Highway Research Centre in McLean, VA established a living laboratory in a work zone on I–95 between Springfield and Lorton, VA. Using an instrumented vehicle, they obtained a large amount of behaviour data of vehicles running through the work zone. Based on this data, they studied the car-following behaviour traffic characteristics in work zone[6]. Wu used an unmanned aerial vehicle(UAV) and recorded the traffic video of a maintenance work zone, then by means of video analysis, the microscopic behaviour data of vehicles in the maintenance work zone was extracted, and based on the data, Wu studied the lane-changing behaviour of vehicles in maintenance work zones[7].

### 2.2 Vehicle behaviour expressing method

Expressing vehicle behaviour of the actual driving process consists of two main tasks: (1) extracting time series periods of interest from the entire time series, i.e., the segmentation; (2) identifying the extracted time series periods, i.e., the identification.

For the segmentation, there are two main ways: (1) the combination of vehicle bus CAN data segmentation and (2) endpoint detection automatic segmentation. As for the combination of vehicle bus CAN data segmentation, the endpoints of vehicle behaviour were marked using the vehicle's other signals from CAN[8-10]. For example, in the literature[9], braking behaviour starts when the brake lights light on, the behaviour ends when the brake lights light off; when the throttle angle is less than a critical value, the acceleration behaviour begins, and when the throttle angle is larger than the critical angle, the acceleration behaviour finishes. In [8], the vehicle behaviour time series is divided by the use of brake pedal, throttle pedal and retarder. As for the endpoint detection automatic segmentation, the periods of interest are extracted by analysing the data characteristics of discrete time segments and setting thresholds. Endpoint detection automatic segmentation methods are widely used in digital signal processing. Common methods are the bi-threshold method, correlation method, variance method, spectral distance method and so on[11-12]. In [13], an endpoint detection algorithm based on a simple moving average (SMA) of the rotational energy is used for automatic segmentation. In [14], Bayesian Multivariate Linear Model and change points were used to achieve automatic segmentation. Since the first method requires more complicated equipment to obtain other signals, the second segmenting method was used in this research. We used a sound signal processing method to implement the automatic segmentation of vehicle behaviour time series.

Identification belongs to the classic classification and clustering problem in machine learning. In [9], researchers compared the performance of unsupervised k-means clustering algorithm and supervised SVM in accelerating, decelerating and turning behaviour recognition, and found that the supervised SVM algorithm to be more robust. In the literature [13], K-Nearest Neighbours (k-NN) classification method was used to identify 12 kinds of vehicle behaviours such as turning right (90°), turning left (90°), "u" turning (180°) and speeding. Another literature[15] used support vector machine

(SVM) for identification and achieved satisfactory results.

## 3 Data acquisition device and method

To obtain enough data for this research, PVMBDADs are adopted. This is a 10cm × 7cm × 4cm portable device with a battery making it run independently. There is one GPS, one tri-axial accelerometer and other units integrated into this small device in order to collect the speed, geographical position, time information, acceleration data of three directions and attitude information. While working, the device collects the GPS data in a frequency of 1Hz and accelerometer data in 20Hz. With these sensors, it is efficient to record the microscopic behaviour data of vehicles.

There is a slot shell designed for the device, which can store a highway toll card. If condition allows, with the cooperation of expressway management departments, we can distribute devices together with toll cards to passing vehicles in the entrance toll stations of expressways and recycle devices and cards in the exit toll stations, then the devices can collect the microscopic behaviour data of vehicles all along the way. As China is now vigorously promoting the expressway composite passing cards [16], they will have a very bright application prospect if such composite cards that can obtain the microscopic behaviour data of vehicles are adopted.

Of course, it can also be applied to a naturalistic driving study (NDS) since this device can operate independently. Just put one device on an experimental vehicle then the vehicle can be transformed into an instrumented vehicle and collect the vehicle data in a naturalistic mode. The external appearance and internal structure of the device are shown in Fig. 3.

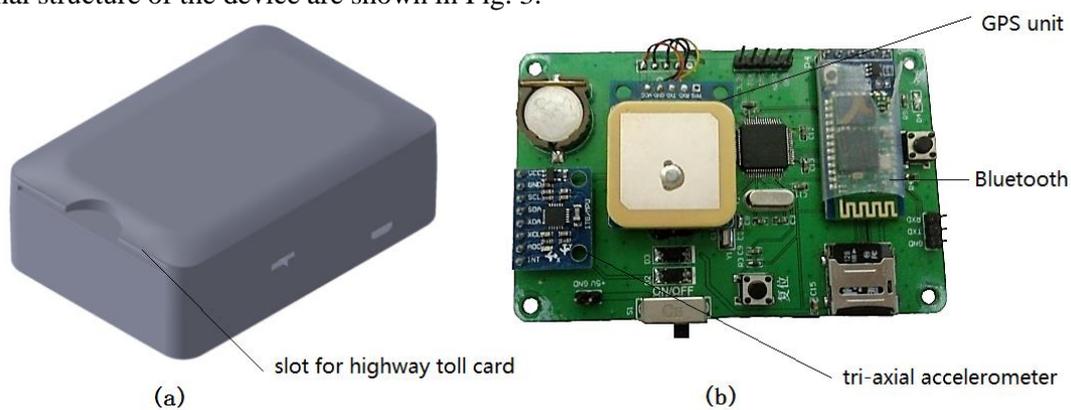

**Fig. 3.** A novel portable data acquisition device

## 4 Vehicle behaviour expressing

The typical microscopic behaviour data collected is as shown in Figure 4 below. The acceleration and speed change over time while the vehicle runs. In Fig. 4, the period where acceleration changes slightly may represent an L&C driving behaviour, while the period with violent acceleration change may mean that the vehicle is taking an accelerating behaviour. The purpose of expressing vehicle behaviour is to identify what behaviour the vehicle is taking in a specific period based on the microscopic data. As introduced in the related work, this task consists of two steps, i.e., segmenting and identifying.

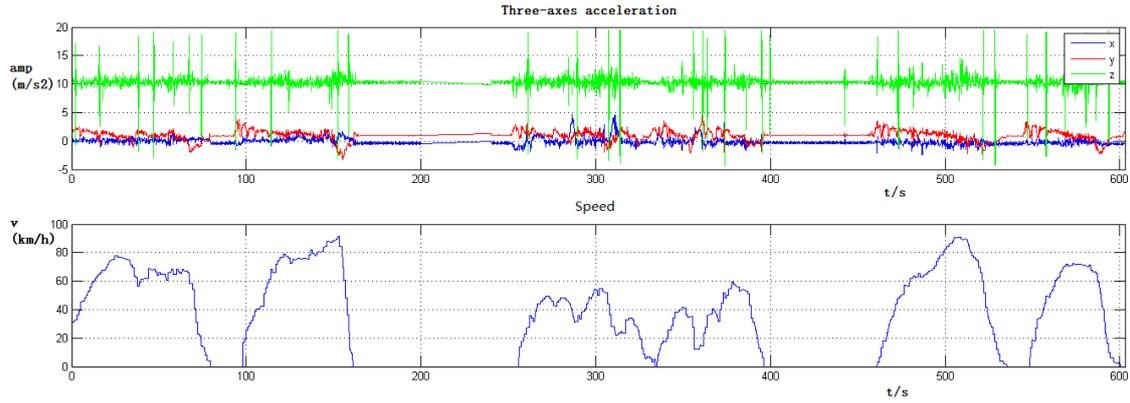

**Fig.4.** A typical microscopic behaviour data sample

### 4.1 Automatic segmentation - endpoint detection technology

For vehicle behaviour, accelerating, decelerating and turning have more significant data features than stopping and L&C. After these special behaviour periods are extracted, the rest are stopping and L&C vehicle behaviour periods, which can be identified in a simple way. Thus, the key point of vehicle behaviour time series segmentation is to extract the accelerating, decelerating and the turning periods. For convenience, this paper defines these key periods as periods of interest (POIs). A short-time energy based one-parameter bi-threshold endpoint detection technology is used to carry out the automatic segmentation and extract out the POIs. Since this method was introduced with detail in paper [17], for more details please refer to the paper. A brief introduction to the main processes is introduced as follows.

Take the longitudinal acceleration ax as an example, as shown in Fig. 5, figure (a) shows the change of ax with time when a vehicle gets close to a maintenance work zone and figure (b) shows the corresponding change of the short-time energy of ax. As is seen from figure (b) obviously, some of the sections fluctuate greatly, which means that the vehicle takes some types of behaviours (such as deceleration) there. To extract out the section with great fluctuation, a higher threshold T2 is set, then the sections with acceleration larger than T2 is of course POIs. A lower threshold T1 is used to find the finer start and end points of this POI. By the endpoint detection method, this section, that is the segment between the green vertical solid line and the red vertical dashed line in figure (a) can be extracted.

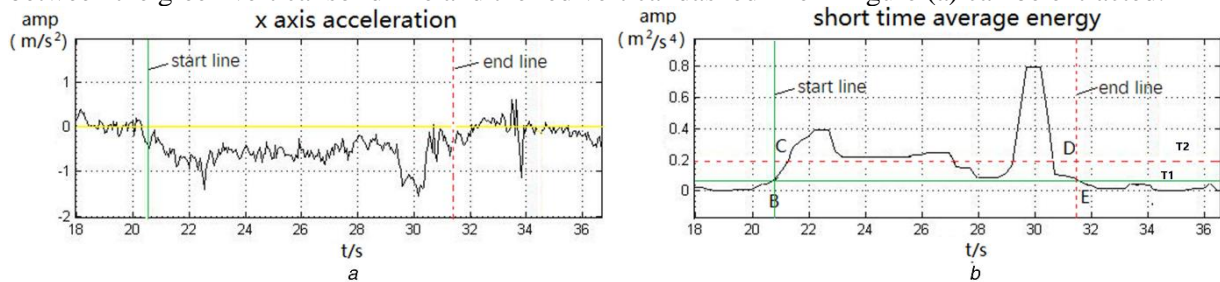

**Fig. 5.** Demonstration of the short-time energy based one-parameter-bi-threshold endpoint detection

The key to the endpoint detection is the setting of two thresholds T1 and T2, i.e., the green horizontal solid line and the red horizontal dashed line on figure (b) respectively. According to different research objectives, there are mainly two methods for determining the values of T1 and T2. The first method is an adaptive method, taking values based on the behaviour data characteristics of each vehicle itself, such as taking 30% of the maximum acceleration of the vehicle behaviour data, or the median, or the average value as the values of T1 and T2. This method is able to shield the difference between vehicles. For example, the change for vehicles of general quality may be very obvious, whereas the acceleration change of the vehicles with high quality may be not obvious. The adaptive method can guarantee the POIs be extracted out regardless of the difference between vehicles. The second method is a determinative method, that is taking determinative values for T1 and T2, for example taking the value of T2 as 1.25m/s2. This method has obvious physical meaning, for example, study shows that when the longitudinal acceleration is larger than 1.25m/s2, the driver or the passengers will feel very uncomfortable [18], thus the vehicle behaviour with acceleration larger than this threshold is riskier to cause accidents and can be defined as unsafe behaviour. Using this method, the sections extracted are unsafe behaviours, which can be used for work zone safety-related researches. What should be noted is that since the short-time energy is used as the segmentation parameter, the determinative threshold value should be transformed to short-time energy.

In this study, the difference between the x-direction and y-direction acceleration of POIs is obvious, but not significant in the z-direction acceleration. Due to the low sampling frequency, other data is not suitable for this automatic segmentation method. Therefore, the automatic segmentation method is only applied to the x-axis acceleration and y-axis acceleration time series.

Automatic segmentation of different time series is performed independently. After the segmentation process, two different sets of schemes will be generated, i.e., the green x-axis acceleration segmentation scheme and the red y-axis acceleration segmentation scheme as shown in Fig. 6. There are overlapping periods and their own unique periods corresponding to different vehicle behaviour of the actual driving process. For example, a period with only a large x-axis acceleration may represent a linear accelerating behaviour, a period with only a large y-axis acceleration may represent a turning with constant speed behaviour, and both may represent a decelerating turning. After POIs are extracted using the automatic segmentation technique mentioned above, the remaining periods are either stopping or L&C vehicle behaviour, which can be identified by speed judgment (the purple scheme in Fig. 6). In the actual processing, we combined the speed into the x, y-axis acceleration segmentation schemes, and achieved the segmentation and recognition of stopping, L&C, linear accelerating, linear decelerating and turning (the bottom scheme in Fig. 6).

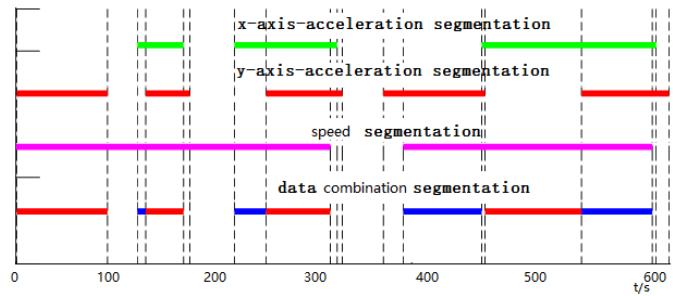

**Fig. 6.** Data combination process

It can be observed that through the automatic segmentation and data combination, the vehicle behaviour time series can be divided into three categories: stopping, L&C and POIs, in which the stopping and L&C vehicle behaviours are determined, while the POIs category is too general. The POIs include eight sub-categories, i.e. L&A, L&D, TL&A, TR&A, TL&C, TR&C, TL&D, and TR&D. So it needs to be further divided by other methods. In this study, the support vector machine (SVM) method is adopted.

### 4.2 SVM based Vehicle Behavior Identification

SVM, as a tool for solving problems in classification, regression, and novelty detection, has been widely involved in various researches. The main idea of SVM is to find a hyperplane as the decision boundary so that the geometric margins between different categories are maximised and the empirical classification error is minimised. SVM is already a very mature method, with good generality, good robustness, high efficiency, simple calculation, and perfect theory [19]. As reviewed in the related work that SVM can achieve high accuracy in behaviour identification, in this research SVM is also used to identify further the POIs extracted by the automatic segmentation technology as L&A, L&D, TL&A, TR&A, TL&C, TR&C, TL&D, and TR&D.

#### 4.2.1 Data acquisition

The establishment of the SVM classification model requires training data and testing data. Measured data was used in this study. Training data was obtained on a specifically selected test road, with excellent pavement condition, a sufficiently long straight line for linear accelerating and decelerating tests, and several different corners for turning tests, as shown in Fig. 7(a). Test data collecting was carried out on a 10 km long normal road section of Caoan Highway, as shown in Fig. 7(b).

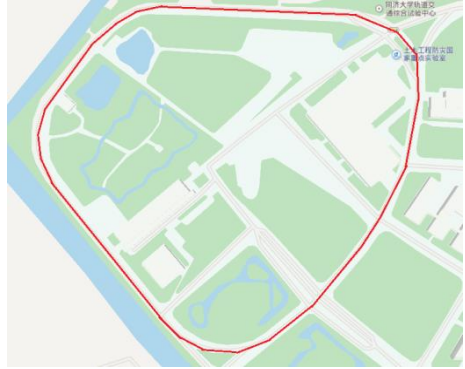

(a) Map of training road

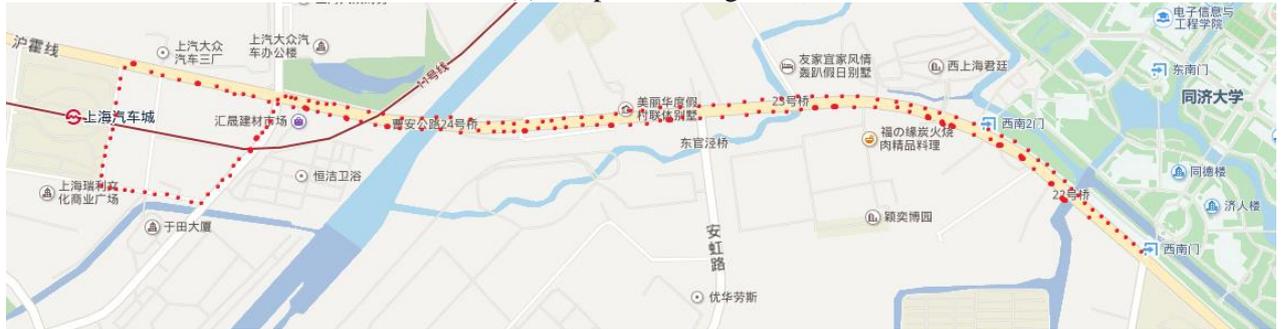

(b) Map of test road
**Fig. 7.** Maps of training and test roads

Before data acquisition, a PVMBDAD was fixed horizontally at the front of the vehicle. The direction of the device was adjusted so that the positive direction of the x-axis acceleration coincides with the driving direction of the vehicle, and the y-axis acceleration direction is consistent with the vehicle's lateral direction, as shown in Fig. 8.

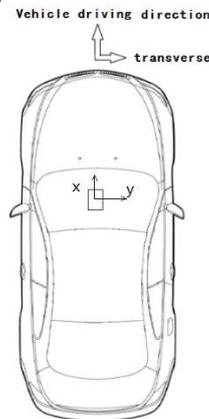

**Fig. 8.** Location and direction of the PVMBDAD in a vehicle

In order to accurately record the start and end points of the vehicle behaviour test series, a dedicated Android application (App) was developed for the device. The App was connected to the device via Bluetooth and can show real-time data (longitude, latitude, three-axis acceleration, speed and time) collected by the device. One can quickly select the type of vehicle behaviour to be recorded, then the start and end points will be written into the database timely and in a standard way. The main interface of the App is shown in Fig. 9.

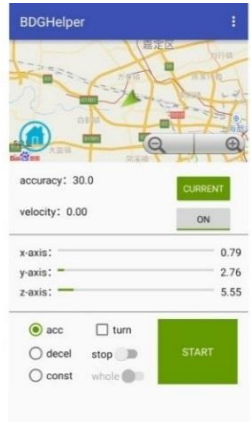

**Fig. 9.** Main interface of the App

To eliminate the impact of other factors, the driver, experiment vehicle, device, and the orientation of the device were consistent in the training and test process. The vehicle travelled on the training road several times and the App was used to keep track of the start and end points of the vehicle's behaviour until a sufficient number of test series periods were collected. Then the vehicle ran in a naturalistic mode on the test road, and behaviours were also recorded with the App for calculating the classification accuracy of the SVM model later. A total of 81 valid periods were collected on training road, including 32 L&A, 13 L&D, 7 TL&A, 6 TL&D, 4 TL&C, 5 TR&A, 12 TR&D, and 2 TR&C.  A total of 27 behaviour periods were obtained on test road, including 13 L&A, 9 L&D, 3 TR&D and 2 TL&A.

*4.2.2 Features*

Through the training data acquisition process, data of 8 kinds of vehicle individual behaviours were collected, that is, L&A, L&D, TL&A, TR&A, TL&C, TR&C, TL&D, and TR&D. Based on these data, the accelerations in the x and y directions line chart and the speed line chart are plotted respectively in Fig. 10. It can be seen that: (1) When the vehicle is moving linearly, the y-direction acceleration fluctuates smoothly near zero; (2) When turning, the acceleration in the y-direction fluctuates greatly; (3) The difference between accelerating, driving with constant speed and decelerating are the trends of the speed curve and the overall value of the x-axis acceleration.

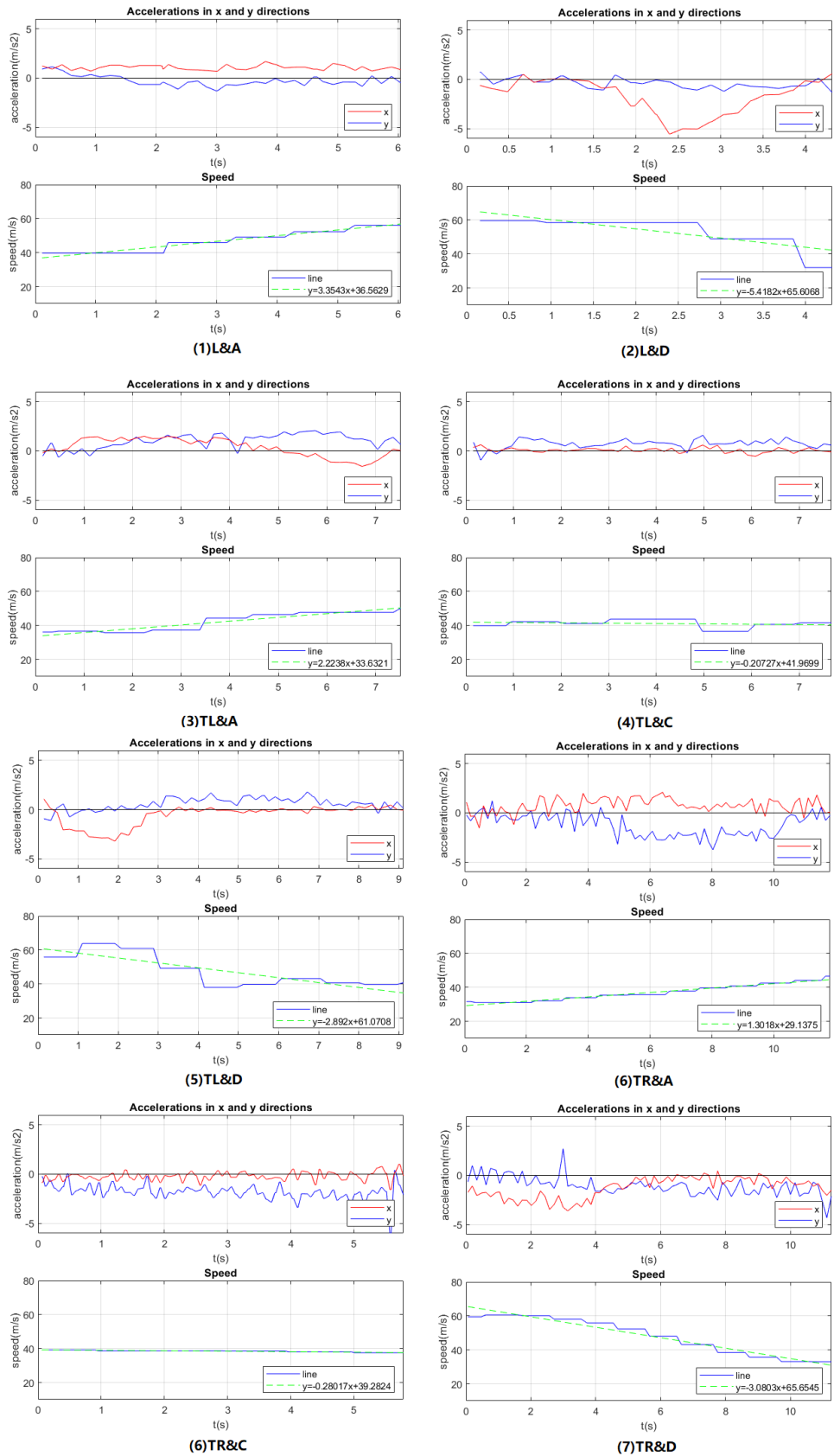

Fig. 10. Typical data line charts of eight periods of interests (POIs)

Based on the above analysis and many practical verifications, the following features are selected: (1) speed gradient (k), (2) mean of x-axis acceleration (x_mean), (3) standard deviation of x-axis acceleration (x_stev), (4) mean of y-axis acceleration, (5) standard deviation of y-axis acceleration (y_stev).

### 4.2.3 Training and Verification

There are many off-the-shell SVM toolkits available, and the *libsvm* package is one of the most famous packages [20]. It has relatively few parameters to be tuned for SVM and provides many default parameters and cross-validation method for parameters optimisation, with which many problems can be solved well. *libsvm* provides a number of kernel functions, including the linear function, polynomial function, radial basis function (RBF) and a sigmoid kernel function. RBF is used in this paper, which has two parameters C and $\gamma$ to be tuned. In this paper, the cross-validation and grid search techniques are used to obtain the optimal values of the parameters. The data analysis processes in this paper were operated on Matlab. For detailed algorithm please refer to[20].

All the 81 groups of training data were chosen as the training data set and the 27 periods of test data were chosen as the test data set. Using the normalized function mapminmax of Matlab, the training data set was normalized to the [0,1] interval[21], then using cross-validation and grid search method, it is found that when C = 1, $\gamma$ = 4, the best accuracy can be achieved as 90.1%, thus verified that the SVM model works well in behaviour identification.

# 5 Vehicle behaviour distribution displaying

## 5.1 Distribution displaying method

Now that the vehicle behaviour expressing method has been established, for behaviour data time series of any vehicle, by endpoint detection, the POIs can be detected, and by data combination and SVM model, the behaviour of any POI can be identified. There are lots of vehicles running on the road network every day, because of the maintenance work zones, vehicles will take common behaviours and show some patterns in space. Analysing the behaviour data of all vehicles running through work zones and drawing the behaviours on maps, we can get a behaviour distribution of the work zones. However, simply adding all the behaviours of all vehicles on maps cannot show us the real traffic characteristics of maintenance work zones, because the existence of both intra-driver heterogeneity and inter-driver heterogeneity within microscopic data [6], i.e., a driver exhibits inconsistent driving behaviour when exposed to different driving environments, and different drivers exposed to the same driving environment may behave differently. Here, we use kernel density analysis to cluster periods with similar behaviours on the map, which clearly shows us the general behaviour distribution in work zones. If there is a high-density clustering centre, it means that most vehicles will take a similar behaviour at that location. For example, if there is an L&D clustering centre in front of the work zone, it means that most vehicles will decelerate at that location.

Kernel density analysis calculates the density of point features around each output raster cell. Conceptually, a smoothly curved surface is fitted over each point. The value is highest at the location of the point and diminishes with increasing distance from the point, reaching zero at the search radius distance from the point. The density at each output raster cell is calculated by adding the values of all the kernel surfaces where they overlay the raster cell centre. The kernel function is based on the quadratic kernel function described in Silverman (1986, p. 76, equation 4.5) [17, 22, 23].

To test the feasibility and performance of methods put forward above, we conducted one experiment and collected the vehicle behaviour data in a maintenance work zone. The work zone was set on the S20 expressway in Shanghai. S20 has a total of 8 lanes in two directions. The task of the maintenance is to repair the joint of the first lane. Two rightmost lanes were closed for the work zone. The layout of the work zone is shown in Fig. 11.

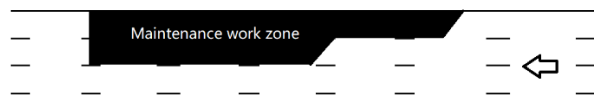

**Fig. 11.** The layout of the maintenance work zone

For the data collecting process, two private cars were recruited. Each car was instrumented with a PVMBDAD and ran through the maintenance work zone for five times. Then segmenting and behaviour identifying processes were carried out on the microscopic data. At last, use kernel density analysis to cluster periods with the same behaviour type. The kernel density analysis is a common tool in Geography Information System(GIS) software. In this research, ArcGIS is used to carry out the analysis.

To better show the behaviour distribution of vehicles, we adopted two displaying strategies, i.e., using different legends and using a unified legend as is shown in Fig. 12(a) and Fig. 12(b), respectively. The figures in the legends are all kernel density values, which indicates the percentage of vehicles taking the same behaviour. Carrying out the same kernel density analysis to the extracted behaviour segments of a road section where 100% vehicles take the same behaviour (e.g., on a flat and straight road section with good traffic condition, all vehicles take the L&C behaviour), a maximum kernel density of $10.94 \times 10^9$ can be got. Then other kernel density values can also be transformed into vehicle percentage by Formula 1. The kernel density values and the corresponding vehicle percentages used in the unified legend are shown in Table 1. For example, when the kernel density value takes $4.5 \times 10^9$, the corresponding vehicle percentage is 41.1%, that means for ten vehicles running through that position, nearly four vehicles will take the same behaviour there.

$$percentage = \frac{kerneldensity}{4.5 \times 10^9} \times 100 \qquad \text{Formula 1}$$

Table 1 Kernel density and the corresponding vehicle percentage

| Kernel density($\times 10^9$) | 0.5 | 1 | 1.5 | 2 | 2.5 | 3 | 3.5 | 4 | 4.5 |
|---|---|---|---|---|---|---|---|---|---|
| Vehicle percentage(%) | 4.6 | 9.1 | 13.7 | 18.3 | 22.9 | 27.4 | 32.0 | 36.6 | 41.1 |

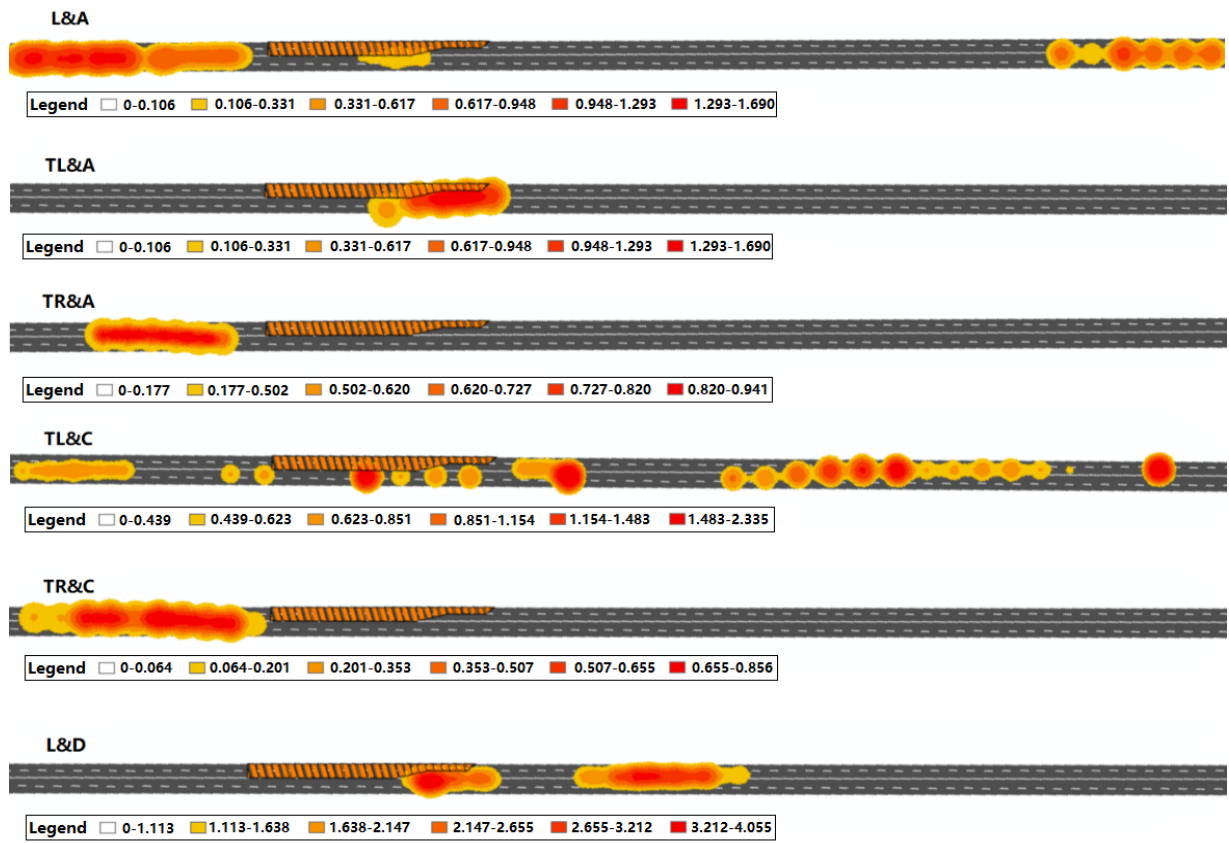

(a)Different legends (kernel density$\times 10^9$)

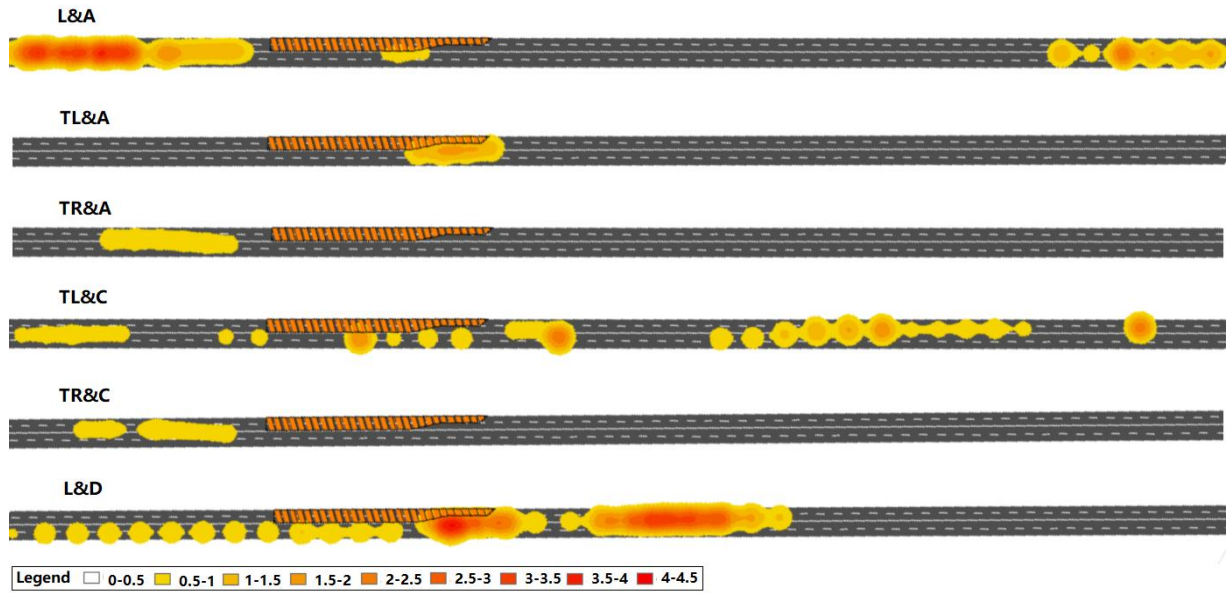

(b)Unified legend (kernel density $\times 10^9$)
**Fig. 12** Behaviour distributions of vehicles

Comparing the maps above, it is very intuitive to see the distributions of different vehicle behaviours in the maintenance work zone. The following conclusions can be drawn:

(1) When different legends are used, the main purpose is to show their own unique distribution characteristics of different vehicle behaviours. It can be clearly seen that there are mainly six types of vehicle behaviours when the vehicles pass through the maintenance work zone, i.e., L&A, TR&A, TL&C, TR&C, L&D and L&A. The locations of distributions of various vehicle behaviours in the maintenance work zone are also very obviously displayed on the map, for example, the TL&A behaviour is mostly distributed in the upstream transition area, the TR&A behaviour is mainly distributed in the termination area, and the L&D behaviour is primarily distributed in the warning area and the upstream transition area. This displaying method can examine the effects of traffic control, such as speed limit control and traffic channelisation. As is seen from the distribution map of L&D that vehicles only decelerate until they are close to the maintenance work zone, so the variable speed limit effect is not obvious. As for the traffic channelisation, if the total number of types of vehicle behaviours is not changed after the channelisation measures are taken, it may indicate that the channelisation effect is not satisfactory, or if the number is found to be significantly reduced, it means that the channelisation effect is very good. Finally, the distribution of vehicle behaviour can show us how vehicles actually travel in the maintenance work zone, whether the distributions are consistent with our previous assumptions to provide a reference for the improvement work of maintenance work zone design.

(2) When a unified legend is used, it is mainly to compare the proportions of different vehicle behaviour distributions and intuitively display them. It can be seen that the L&A and L&D are the main behaviours in the six types of vehicle behaviours, followed by TL&A behaviour and TL&C behaviour. Others such as the TR&A and TR&C behaviours are relatively less distributed. The proportions tell us what should be paid attention to. For example, if some channelisation measures are taken to reduce the less-distributed behaviours, the traffic flow will be more consistent and the travelling efficiency will be improved.

### 5.2 Discussion

The following issues are further discussed as follows:

(1)As can be seen from the above figures, this method is an effective complement to existing traffic characteristic analysis methods. Existing methods can only describe the macroscopic traffic characteristics in a few discrete cross-sections of the road, but the method of this paper can obtain a continuous distribution map, and show how the vehicles operate in the maintenance work zone clearly at a glance. Though the distribution map got by kernel density analysis to individual behaviours is also a kind of macroscopic traffic characteristic, it is fundamentally different from the existing macroscopic traffic characteristics analysis. Because the method in this paper carries out a detailed analysis to each vehicle individual first and then obtains the common pattern exhibited by all individuals, whereas the existing analysis methods lack the step of detailed analysis to the individuals, so there is no way to obtain the above-mentioned continuous and fine effect.

(2)Using the kernel density analysis, the common distribution pattern of the vehicle behaviours in space can be displayed. The advantage of the kernel density analysis is that it can reflect the vast majority of cases, and shield out individual outliers. Therefore, even if there is an identification error happened in the SVM behaviour identification process, or the vehicle experienced an abnormal situation, it will not affect the final distribution. Therefore, the method for expressing and displaying the behaviour distribution put forward in this paper is most suitable for studying the influence of maintenance work zones on the behaviour distribution, analysing the distribution pattern and the distribution change regulation and then proposing new safety assessment method and improved design concept.

(3)The functions of using different legends and unified legend are shown in Section 5.1. In summary, using different legends, each behaviour distribution can be displayed in detail, which is mainly used to analyse the spatial distributions of vehicle behaviours; Using the unified legend, the proportions of different vehicle behaviour distributions can be compared to show the factors that should be paid attention to in maintenance work zones.

(4) Endpoint detection is the key of this method, especially the setting of two thresholds can achieve different results. As introduced in the vehicle behaviour expressing method, there are mainly two methods for determining the values of T1 and T2, the adaptive method and the determinative method. For the adaptive method, it can extract all the POIs out regardless of the differences between vehicles, while using different legends, all the behaviours distribution can be displayed on maps with details. As for the determinative method, unsafe behaviors can be defined and extracted, combined with the unified legend, which can be used to study the unsafe behavior distribution characteristics, especially to study the impacts of factors of maintenance work zones on the distribution of unsafe behaviors and propose novel method for safety assessment for maintenance work zones to correct safety hazards and improve maintenance work zone design concepts and methods.

(5)An actual data collection in a maintenance work zone was carried out in this research and two instrumented vehicles were recruited, which is enough for just demonstrating the method proposed in this paper. However, in order to achieve more accurate distributions in practice, it is best to acquire all or most of the microscopic behaviour data of vehicles travelling through maintenance work zones. If condition permits, making use of the PVMBDADs together with highway toll cards is the better solution. If not, try to recruit more vehicles with PVMBDADs instrumented and run through the maintenance work zone for more times, and a satisfactory accuracy of the distribution results can also be achieved.

## 6 Summary and prospect

By means of portable vehicle microscopic behaviour data acquisition devices, the microscopic behaviour data of vehicles running through work zones was collected. An endpoint detection technology was used to segment the behaviour time series dynamically. Then the corresponding behaviour type of each segment was identified by a SVM algorithm. Finally, kernel density analysis was carried out to cluster similar behaviours and show the general distribution of behaviours in maintenance work zones. Practice verified that the SVM identification algorithm could achieve high accuracy, combined with the dynamic segmentation, it can express the vehicle behaviour accurately. The kernel density analysis works well in clustering similar behaviours and shows the general distribution of vehicle behaviours in maintenance work zones. The behaviour distribution maps can intuitively reflect the traffic characteristics and show the effect of traffic control in maintenance work zones.

Next, we will further define and extract unsafe behaviours, study how maintenance work zones affect the unsafe behaviour distributions exactly and make clear their distribution characteristics and distribution change regulations, which will shed light to novel maintenance work zone safety assessment methods.

## Acknowledgements

This work was sponsored by National Nature Science Foundation of China (Grant NO: 51778482).

## References

[1]Polus, Abishai & Shwartzman, Yechezkel: 'Flow Characteristics at Freeway Work Zones and Increased Deterrent Zones'. Transportation Research Record, 1999, 1657. 18-23. 10.3141/1657-03.
[2]Migletz, J., Graham, J. L., Anderson, I. B., Harwood, D. W., & Bauer, K. M.: 'Work Zone Speed Limit Procedure'. Transportation Research Record, 1999, 1657(1), 24–30. https://doi.org/10.3141/1657-04
[3]A.T. Vemuri, M.M. Polycarpou, P.D. Pant: 'Short-term forecasting of traffic delays in highway construction zones using on-line approximators'. Mathematical and Computer Modelling, 1998, Volume 27, Issues 9–11, Pages 311-322. https://doi.org/10.1016/S0895-7177(98)00066-1.
[4]Benekohal, Rahim F., Ahmed-Zameem Kaja-Mohideen, and Madhav V. Chitturi: 'Evaluation of construction work zone operational issues: Capacity, queue, and delay'. ITRC FR 00/01-4, Illinois Transportation Research Center, Champaign, IL 2003.
[5]Lochrane, Taylor, Al-Deek, Haitham, Paracha, Jawad, et al.: 'Understanding Driver Behavior in Work Zones'. Public Roads, 2013, 76.

[6]Berthaume, A. L., James, R. M., Hammit, et al.: 'Variations in driver behavior: an analysis of car-following behavior heterogeneity as a function of road type and traffic condition'. Transportation research record, 2018, 0361198118798713.
[7]Jiangling, WU: 'Lane-changing Behavior Analysis, and Modeling of lane-changing in Work Zones on Freeways'. (Doctoral dissertation). Chang'an University, Xi'an, China, 2017.
[8]D'Agostino, Claire, et al.:'Learning-Based Driving Events Recognition and Its Application to Digital Roads.' IEEE Transactions on Intelligent Transportation Systems, 2015, 16.4(2015):2155-2166.
[9]Ly, M. Van, S. Martin, and M. M. Trivedi: 'Driver classification and driving style recognition using inertial sensors'. Intelligent Vehicles Symposium IEEE, 2013:1040-1045.
[10]Klauer, S G, et al. : 'The Impact of Driver Inattention on Near-Crash/Crash Risk: An Analysis Using the 100-Car Naturalistic Driving Study Data'. U.s.department of Transportation Washington D.c (2006).
[11]Qu Yuan, Peng Xuan, Wang Bingxi: 'Practical Speech Recognition Basics'. National Defense Industry Press, 2005, China
[12]Lyons, Richard G.: 'Understanding Digital Signal Processing'. Pearson Education, 2010.
[13]Johnson, D. A, and M. M. Trivedi: 'Driving style recognition using a smartphone as a sensor platform'. International IEEE Conference on Intelligent Transportation Systems IEEE, 2011:1609-1615.
[14]Bender, Asher, et al. : 'An Unsupervised Approach for Inferring Driver Behavior From Naturalistic Driving Data'. IEEE Transactions on Intelligent Transportation Systems, 2015, 16.6(2015):3325-3336.
[15]Fang-fang Zhang: 'Study on a Computer Vision System for Detecting Traffic Conflict between Vehicles at Intersections'. Tongji University,2008.
[16]http://www.gov.cn/xinwen/2018-12/25/content_5351962.htm?_zbs_baidu_bk
[17]Yang, Q., & Xu, Z. : ' Novel Fast Safety Assessment Method for the Buffer Section of Maintenance Work Zone'. IET Intelligent Transport Systems, 2018, DOI: 10.1049/iet-its.2018.5173.
[18]Tianxing Chen. (2014). Study on the evaluation of highway alignment consistency based on acceleration. (Doctoral dissertation, Chang'an University).
[19]Harrington, P. (2012). Machine learning in action (Vol. 5). Greenwich: Manning.
[20]Chang, Chih Chung, and C. J. Lin: 'LIBSVM: A library for support vector machines'. ACM, 2011.
[21]Matlab Chinese Forum. (2010) . 30 Analysis Cases of Neural Network for Matlab.
[22]Silverman, B. W.: 'Density Estimation for Statistics and Data Analysis'.New York: Chapman and Hall, 1986.
[23]Esri. 'How Kernel Density works'. http://resources.arcgis.com/en/help/main/10.1/index.html#//009z00000011000000. Accessed July 21, 2017